\documentclass[aps,prb,twocolumn,floatfix,showpacs,showkeys]{revtex4}
\usepackage{amsmath}
\usepackage{amssymb}
\usepackage{amsfonts}

\begin{document}
\title{Anisotropic electron $g$-factor in quantum dots with spin-orbit interaction}
\author{C. F. Destefani}
\author{Sergio E. Ulloa}
\affiliation{Department of Physics and Astronomy, Ohio University, Athens, Ohio 45701-2979}
\date{\today}

\begin{abstract}
$g$-factor tuning of electrons in quantum dots is studied as
function of in-plane and perpendicular magnetic fields for
different confinements. Rashba and Dresselhaus effects are
considered, and comparison is made between wide- and narrow-gap
materials. The interplay between magnetic fields and intrinsic
spin-orbit coupling is analyzed, with two distinct phases found in
the spectrum for GaAs in perpendicular field. The anisotropy of
the $g$-factor is reported, and good agreement with available
experimental findings is obtained.
\end{abstract}

\maketitle

Spin properties in semiconductor quantum dots (QDs) have become a
field of intense research because of the possible use of the
electron spin degree of freedom as a quantum bit.\cite{vicenzo} It
is then essential to have a clear understanding of the processes
that may induce spin relaxation on electrons in the QD; long spin
relaxation times, and as pure as possible spin states, are
required so that spin can indeed transport information without
losses. Insights on the purity of the spin degree of freedom of
electrons in QDs can also be extracted from measurements of their effective $%
g$-factor.

Manipulation of $g$-factor in semiconductors and description of
its tensorial nature have been considered theoretically as well as
experimentally. Among various techniques, appropriate system
design has achieved gate-voltage control of $g$-factor of
electrons in a quantum well\cite{sallis} and spin manipulation
using gigahertz electric fields.\cite{kato} $g$-factor
measurements have been reported in QDs by means of
capacitance\cite{medeiros} and energy\cite{hanson,lindemann}
spectroscopies, for example. The tensorial nature of electron
$g$-factor in spherical QDs, \cite{kiselev} as well as
surface\cite{rodina} and spatial confinement\cite{silvio} effects,
where spin-orbit (SO) coupling plays an important role, have been
addressed theoretically. We have recently reported on the
influence of SO coupling on the electronic spectrum of 2D
parabolic QDs.\cite{dest} Spin-orbit effects on $g$-factor have
been addressed for 2D electrons in a quantum well\cite{raef} and
for electrons in a parabolic QD,\cite{rog} with emphasis on the
difference between narrow- and wide-gap compounds.

One of the main causes of spin relaxation and $g$-factor variation
is the SO interaction. When QDs are built in semiconductors of
zincblende structure in the plane of a 2D system, there are two
possible forms of SO coupling, namely the Rashba\cite{rashba} and
Dresselhaus\cite{dres} interactions; the former is due to the
surface inversion asymmetry (SIA) induced by the 2D confinement,
while the latter is caused by the bulk inversion asymmetry (BIA)
intrinsic in zincblende structures. The SO coupling mixes spins
with different orientations in the Zeeman sublevels, which yields
an \textit{intrinsic} spin relaxation source and produces
variations in the QD $g$-factor from the pure Zeeman splitting
expected in a magnetic field.

In this work we study the anisotropy of effective $g$-factors in
2D parabolic QDs, and analyze the intricate competition\cite
{SOZeeman} between external magnetic fields and intrinsic SO
couplings. Wide- and narrow-gap materials with different
confinement potentials are considered under in-plane and
perpendicular magnetic fields. We show that the $g$-factor can be
tuned to have positive, zero, and negative values at given
perpendicular fields, and find that two distinct phases can be
present in the QD spectrum, where at low (high) fields the SO
coupling enhances (suppresses) the Zeeman sublevel splitting. We
find that the widely used perturbative approach\cite{gover} based
on a unitary transformation\cite{aleiner} of the SO Hamiltonian
has strict limitations in QDs if the well defining the 2D
confinement has small width. We also show that even for
\textit{in-plane} magnetic fields -- typically assumed to yield
the bulk $g$-factor without orbital contributions -- the SO
coupling reduces the value of $g$, especially for large QDs.

The QD is defined by an in-plane parabolic confinement, $%
V(\rho )=m\omega _{0}^{2}\rho ^{2}/2$, where $m$ ($\omega _{0}$)
is the electronic effective mass (confinement frequency), while
the perpendicular confinement $V(z)$ is assumed strong enough to
reduce the electronic states to the first conduction subband of
the quantum well; its function is $\varphi (z)=\sqrt{2/z_{0}}\sin%
\left( \pi z/z_{0}\right) $, $z_{0}$ being the QD vertical width
for hard wall potential. In the absence of SO interactions, the QD
Hamiltonian is $H_{0}=\hbar
^{2}\mathbf{k}^{2}/2m+V(\rho )+g_{0}\mu _{B}%
\mathbf{B}_{\perp }\cdot \mathbf{\sigma }/2$, where
$\mathbf{k}=-i\mathbf{\nabla }+e\mathbf{A}/(\hbar c)$ is the
canonical momentum, $g_{0}$ is the material bulk $g$-factor,
$\mathbf{\sigma }$ stands for the Pauli matrices, $\mu _{B}$ is
the Bohr magneton, and $\mathbf{A}=B_{\perp }\rho (-\sin \phi
,\cos \phi ,0)/2$ describes a perpendicular magnetic field $\mathbf{B}%
_{\perp }=B_{\perp }(0,0,1)$, which lifts both orbital and spin degeneracies
of levels. $H_{0}$ yields the Fock-Darwin (FD) spectrum,\cite{pawel} $%
E_{nl\sigma _{z}}=(2n+|l|+1)\hbar \Omega +l\hbar \omega _{c}/2+g_{0}\mu
_{B}B_{\perp }\sigma _{z}/2$, with effective (cyclotron) frequency $\Omega =%
\sqrt{\omega _{0}^{2}+\omega _{c}^{2}/4}$ ($\omega _{c}=eB_{\perp
}/mc$), where $n=0,1,2,...$ and $l=0,\pm 1,\pm 2,...$ are
respectively the radial and azimuthal quantum numbers. The FD
eigenfunctions are $\Psi _{nl}(x,\phi )=R_{n|l|}(x)e^{il\phi%
}/\sqrt{2\pi }$, where $R_{n|l|}(x)=\sqrt{2n!/(n+|l|)!}/\lambda%
x^{|l|}e^{-x^{2}/2}L_{n}^{|l|}(x^{2})$, in terms of associated
Laguerre polynomials $L_{n}^{|l|}$, $\lambda =\sqrt{\hbar
/(m\Omega )}$ is the effective QD lateral length, and $x=\rho
/\lambda $.

The SO Hamiltonian added to $H_{0}$ is $H_{SO}=H_{SIA}+H_{BIA}$. The SIA
term, due to the full confining potential $V(\mathbf{r})=V(\rho )+V(z)$ and
with coupling parameter $\alpha $, is given by $H_{SIA}=\alpha \mathbf{%
\sigma }\cdot \mathbf{\nabla }V(\mathbf{r})\times \mathbf{k}$. It can be
separated as $H_{SIA}=H_{R}+H_{LAT}$, where $H_{R}=-\alpha /\lambda
(dV/dz)[\sigma _{+}L_{-}A_{-}+\sigma _{-}L_{+}A_{+}]$ is the Rashba term\cite%
{rashba} caused by the interfacial electric field generated by the
perpendicular confinement, and $H_{LAT}=\alpha (\hbar \omega
_{0}/l_{0}^{2})\sigma _{z}[L_{z}+(\lambda /l_{B})^{2}x^{2}/2]$ is
due to the lateral confinement (diagonal in the FD basis); we
define $L_{\pm }=e^{\pm i\phi }$, $\sigma _{\pm }=(\sigma _{x}\pm
i\sigma _{y})/2$, $L_{z}=-i\partial /\partial \phi $, and $A_{\pm
}=\mp \partial /\partial x+L_{z}/x+(\lambda /l_{B})^{2}x/2$, and
the confining (magnetic) length is $l_{0}=\sqrt{\hbar /(m\omega
_{0})}$ ($l_{B}=\sqrt{\hbar /(m\omega
_{c})}$). The $z$-confinement yields the BIA term,\cite{dres} $%
H_{BIA}=\gamma \lbrack \sigma _{x}k_{x}\left(
k_{y}^{2}-k_{z}^{2}\right) +\sigma _{y}k_{y}\left(
k_{z}^{2}-k_{x}^{2}\right) +\sigma _{z}k_{z}\left(
k_{x}^{2}-k_{y}^{2}\right) ]\equiv H_{D}^{L}+H_{D}^{C}$;
the linear contribution is $H_{D}^{L}=-i\gamma \left\langle
k_{z}^{2}\right\rangle /\lambda \left[ \sigma
_{+}L_{+}A_{+}-\sigma _{-}L_{-}A_{-}\right] $, while the cubic is $%
H_{D}^{C}=i\gamma /\lambda ^{3}[\sigma _{-}L_{+}^{3}H_{1}+\sigma
_{+}L_{-}^{3}H_{2}+\sigma _{-}L_{-}H_{3}+\sigma _{+}L_{+}H_{4}]$, where $%
H_{i}=A_{i}+(\lambda /l_{B})^{2}B_{i}+(\lambda
/l_{B})^{4}C_{i}+(\lambda /l_{B})^{6}D_{i}$, with $i=1,2,3,4$. The
sixteen functions $A_{i}$, $B_{i}$, $C_{i}$, $D_{i}$ are
known,\cite{dest} $\gamma $ is the coupling parameter, and
$\left\langle k_{z}^{2}\right\rangle =(\pi /z_{0})^{2}$ for hard
wall confinement. The total QD Hamiltonian,
$H=H_{0}+H_{LAT}+H_{R}+H_{D}^{L}+H_{D}^{C}$, is diagonalized in a
basis set including $110$ FD states.

When considering an {\em in-plane} field ($B_\parallel$) we take the full
Hamiltonian above at zero $B_\perp$-field, $H(B_\perp =0)$, plus the Zeeman
contribution $g_{0}\mu _{B}\mathbf{B}_{\parallel }\cdot
\mathbf{\sigma }/2$, which lifts only the spin degeneracy of
levels.

Before discussing the QD effective $g$-factor, we comment on the
influence of SO coupling on the spectrum and its competition with
external magnetic fields. This is done in Figs.\ \ref{f1} and
\ref{f2} respectively for InSb\cite{parin} and GaAs\cite{parga}
QDs, the former having larger SO coupling and Zeeman splitting
than the latter. In these two figures, left (right) panels refer
to an in-plane (perpendicular) field, and dotted lines in panels
$A$ and $B$ refer to the pure FD spectrum (no SO-coupling). In
panels $A$ and $B$ of Fig.\ \ref{f1}, the main effects of the SO
interaction on the InSb QD levels are as follows: crossings
between FD levels are converted into anticrossings (ACs) according
to the selection rules\cite{dest} of $H_{SO}$ (and shifted to
higher field values in the perpendicular case); the low-field
spectrum is displaced to lower energies, while at high-field is
not altered much; zero-field splittings appear in the spectrum,
and the original FD energy shells are separated into components
according to the total angular momentum projection, $j=l+\sigma%
_{z}/2$. Panels $C$ and $D$ ($E$ and $F$) of Fig.\ \ref{f1} show
the field-evolution of spin $\left\langle \sigma _{z}\right\rangle%
$ (orbital $\left\langle l\right\rangle $) angular momentum
expectation values for the seven lowest QD levels, where it is
clearly visible how the SO coupling mixes states, resulting in
$\sigma _{z}$ and $l$ no longer reflecting pure FD levels,
especially around AC points. For the purposes of discussion, we
still label states by the numbers $\{l,\sigma _{z}\}$ even under
SO interaction.

\begin{figure}
\caption{InSb QD spectrum [\onlinecite{parin}]. Left (right)
panels refer to in-plane (perpendicular) magnetic field. Panels
$A$ and $B$ show spectra (dotted lines refer to QD without SO
interaction). Panels $C$ and $D$ ($E$ and $F$) show
field-evolution of spin (orbital) angular momentum expectation
values for the seven lowest levels; ground state is spin-up.
Labels in $C$ and $D$ show which SO term is responsible for the
lowest ACs. Colors indicate different levels: black, red, green,
... stand for ground, first excited, second excited, ... states,
in all panels. A sudden change in line color indicates a level
crossing.} \label{f1}
\end{figure}

By comparing left (right) panels in Fig.\ \ref{f1}, one can obtain
the QD level sequence and realize which SO mechanism is
responsible for the level ACs under $B_{\parallel }$ ($B_{\perp
}$). Focusing on the lowest energy levels, notice at $B_{\parallel
}\simeq 0.1$ T that the lowest two states in the first shell have
$\left\vert j\right\vert =1/2$ ($\{l,\sigma_z\}=\{0,1\}$,
$\{0,-1\}$); in the second shell, the two lower states have
$\left\vert j\right\vert =1/2$ ($\{-1,1\}$, $\{1,-1\}$), while the
two upper ones have $\left\vert j\right\vert =3/2$ ($\{1,1\}$,
$\{-1,-1\}$). At $B_{\perp }\simeq 0.1$ T, the only difference in
such sequence is in the ordering of the two upper
states in the second shell, which become $\left\vert j\right\vert =3/2$ ($%
\{-1,-1\}$, $\{1,1\}$). This ordering reversal is due to the influence of $%
B_{\perp }$ on the orbital features of the state, which changes the effect
of SO coupling on these levels; similar features appear in higher shell
states. Even though the ground state does not exhibit AC for any
field-direction, it is also not spin-pure at low magnetic fields; observe
that $\left\langle \sigma _{z}\right\rangle \simeq 0.75$ in both $B_{\perp }$
and $B_{\parallel }$, and it is positive since $g_{0}$ is negative.
Regarding level ACs, the lowest under $B_{\parallel }$ ($\simeq 4.8$ T)
involves the second and fourth QD states in panel $A$, which at that field
are the levels $\{0,-1\}$ and $\{1,1\}$; such levels are connected by the
operators $\sigma _{\pm }L_{\pm }$, and although they appear in both $%
H_{D}^{C}$ and $H_{D}^{L}$ terms, this AC is mostly due to the \textit{%
linear Dresselhaus} term as indicated in panel $C$. On the other hand, the
lowest AC under $B_{\perp }$ ($\simeq 3.3$ T) involves the second and third
QD states in panel $B$, which at that field are the levels $\{0,-1\}$ and $%
\{-1,1\}$; such levels are connected by the operators $\sigma _{\pm }L_{\mp
} $, so that this AC is due to the \textit{Rashba}\ term $H_{R}$, as
indicated in panel $D$. Higher energy ACs are evident at similar values of $%
B_{\parallel }$ (or $B_{\perp }$) and are due to the same SO
mechanism. The different SO terms in $H$, as well as the level
dispersion in the two field-configurations producing the ACs,
result also in different spin mixings. {\em This clearly affects
the QD effective $g$-factor}, as we will see below. ACs due to the
cubic Dresselhaus term are also observed at higher fields: at
$B_{\perp }\simeq 6$ T the lowest AC involves the fourth and fifth
levels, $\{0,-1\}$ and $\{-3,1\}$, connected by the operators
$\sigma _{\mp }L_{\pm }^{3}$. Interestingly, at $B_{\parallel
}\simeq 14$ T, where ACs due to $H_{D}^{C}$ involve higher energy
levels, $\left\langle l\right\rangle $ for all QD levels collapse
to zero, indicating strong orbital mixing with full spin
polarization.

\begin{figure}
\caption{Level spectrum for GaAs QD [\onlinecite{parga}].
$B_{\perp } $-scale of panel $B$ is different from panels $D$ and
$F$. No AC occurs under $B_{\parallel }$. Ground state at
low $B_{\perp } $-fields is spin-down, and it is flipped back to spin-up around
$13.7$ T, as shown in panel $D$.} \label{f2}
\end{figure}

Figure \ref{f2} has the same analysis for a GaAs QD. The left panels show
that in an in-plane field the SO coupling is not strong enough to induce ACs
in the spectrum (panel $A$), \textit{even though zero-field splittings are
clearly observed} for different $\left\vert j\right\vert $ values (e.g., look at the
second shell). Notice that levels present the same ordering as in the low-$%
B_{\parallel }$ InSb QD, and that spin mixing is present in the
spectrum, as states do {\em not} have $\left\langle \sigma
_{z}\right\rangle =\pm 1$ (panel $C$); e.g., the ground state has
$\left\langle \sigma _{z}\right\rangle \simeq 0.75$ at small
fields. This shows the intricate interplay between external fields
and SO coupling in the definition of electronic properties of
QDs.\cite{SOZeeman} In a perpendicular field (right panels of
Fig.\ \ref{f2}), however, a new feature is observed: the SO
interaction \textit{flips} the spins of the two lowest levels at
low fields, so that the QD ground state becomes a spin-down level.
For this reason, the lowest AC at $B_{\perp }\simeq 2$ T (notice
different $B_{\perp }$-scale in panel $B$ compared to $D$ and $F$)
involving the second and third
levels, $\{0,1\}$ and $\{-1,-1\}$, is due to the \textit{%
linear Dresselhaus} term. Observe in panel $D$ that those two
lowest levels have the same value of $\left\langle \sigma
_{z}\right\rangle \simeq -1$ between $4$ and $8$ T, and that the
ground state \textit{flips back} to its expected spin-up character
of a $g_{0}<0$ QD around $13.7$ T, when a level crossing is
verified in the spectrum (not shown). This result prompts one to
consider two distinct ways of defining the QD effective
$g$-factor, and points out strict limits for the validity of the
widely used perturbative approach\cite{gover} (or unitary
transformation\cite{aleiner}) for dealing with SO effects in QDs,
especially if the well thickness $z_{0}$ defining the
$z$-confinement\cite{well} is small. One has to keep in mind
that these results depend on the QD energy (length) scales: higher $%
E_{0}=\hbar \omega _{0}$ (smaller $l_{0}$) QDs are expected to have weaker
SO effects, since their level spacing becomes larger.

What is the `correct' way of defining the effective electron
$g$-factor in a QD? There are two possible definitions involving
the two lowest Zeeman sublevels for each field-direction, namely,
$g_{\perp ,\parallel }^{E}/g_{0}=\Delta E/(g_{0}\mu _{B}B_{\perp
,\parallel })$ or $g_{\perp ,\parallel }^{\sigma
_{z}}/g_{0}=\left\langle \Delta \sigma _{z}\right\rangle /2$,
where $\Delta E$ ($\left\langle \Delta \sigma _{z}\right\rangle $)
is the energy splitting (spin expectation value difference) of
those levels under $B_{\perp ,\parallel }$. Although the first
definition is used operationally in experiments where $\Delta E$
is measured, the latter is intuitively reasonable since the
$g$-factor is a quantity intrinsically related to the spin value
of those levels.  For no SO interaction, both definitions yield
$g_{\perp }=g_{\parallel }=g_{0}$, so that no anisotropy is
present in the $g$-factor (other than the material
anisotropies).\cite{sallis,medeiros} Figures \ref{f3} (InSb) and
\ref{f4} (GaAs) present the $g$-factor for different QD lateral
sizes $l_{0}$, with left (right) panel for in-plane
(perpendicular) fields, and dotted lines show results without SO
coupling. Panels $A$ and $B$ ($C$ and $D$) use the definition of
$g$ in terms of $\left\langle \Delta \sigma
_{z}\right\rangle $ ($\Delta E$), whose values can be inferred from panels $%
C $ and $D$ of Figs.\ \ref{f1} and \ref{f2} (panels $E$ and $F$ of Figs.\ %
\ref{f3} and \ref{f4}). Notice that for InSb in Fig.\ \ref{f3} both
definitions yield basically the same \textit{low-field} results for the $%
B_{\parallel }$ case (panels $A$ and $C$). The drop in $g_{\parallel }$ is
faster in the $g^{\sigma _{z}}$ curves because the two lowest states have
the same spin at higher fields, while $\Delta E$ reaches a constant value
(panel $E$); dotted lines show the field value where the original FD lowest
crossing is converted into an AC by the SO coupling at a given $E_{0}$.

\begin{figure}
\caption{Electron $g$-factor of InSb QDs with different
confinement energies $E_{0}$ (or lateral sizes $l_{0}$). Left
(right) panels refer to in-plane (perpendicular) field. Panels $A$
and $C$ ($B$ and $D$) show $g_{\parallel }$ ($g_{\perp }$), while
panels $E$ and $F$ show sublevel splitting $\Delta E$ for the two
lowest QD states. Panels $A$ and $B$ ($C$ and $D$) show $g$ as
obtained from the $\left\langle \Delta \protect\sigma
_{z}\right\rangle $
($\Delta E$) calculation; values of $\left\langle \Delta \protect\sigma %
_{z}\right\rangle $ for the $15$ meV QD are obtained from Fig.\
\protect \ref{f1}. Dotted lines in all panels refer to
corresponding QDs without SO coupling. $\Delta E$ in panel $F$ is
multiplied by $2$. Color legend is shown in panel $B$; dotted
lines follow same scheme.} \label{f3}
\end{figure}

In the low-field $B_{\perp }$ case (panels $B$ and $D$), however,
the two $g$-factor definitions yield different values mainly at
weaker confinements, although the drop in $g_{\perp }$ is also
faster in the $g^{\sigma _{z}}$ curves ($\Delta E$ values are
found in panel $F$). Observe that for the smallest confinement
energy (largest SO effect) of $3.0$ meV, a sign change is seen in
$g_{\perp }^{\sigma _{z}}$ around $1$ T, which relates to an
unusual crossing involving the \textit{ground} and first excited
states (compare with panel $B$ of Fig.\ \ref{f1} where this
crossing is absent). In both field-directions, smaller $E_{0}$
yields smaller $g$-factor, which shows that SO coupling provides a
channel to manipulate $g$ in QDs under magnetic fields. At the
same time, a measurement of effective $g$ under $B_{\perp%
,\parallel }$ might give information about the values of SO
constants since, as discussed in Fig.\ \ref{f1}, the lowest AC in
the spectrum of QDs in different field-directions is due to a
distinct SO mechanism. Notice the clear anisotropy in $g^{E}$
(panels $C$ and $D$): the same QD confinement shows $g_{\perp
}^{E}<g_{\parallel }^{E}$, since the mixing with higher orbitals
is stronger
for $B_{\perp }$. If the $g^{\sigma _{z}}$ definition is considered (panels $%
A$ and $B$), such anisotropy is not as remarkable at low fields, although a
sign change is obtained at $E_{0}=3$ meV, only in the $g_{\perp }^{\sigma
_{z}}$ curve. Notice that unlike the case with no SO interaction, even $%
B_{\parallel }$\ reduces the $g$-factor and this reduction can in
fact be strong ($\gtrsim 50$ \% at $l_{0}\simeq 300$ \AA ).

\begin{figure}
\caption{Same as Fig.\ \protect\ref{f3} but for GaAs QDs. $g_{\perp }^{%
\protect\sigma _{z}}$ has inverted sign at low fields indicating that SO
coupling is stronger than the Zeeman splitting ($\left\langle \Delta \protect%
\sigma _{z}\right\rangle $ of the $1.1$ meV QD from Fig.\
\protect\ref{f2}). $g_{\perp }^{E}$ has large values at low
fields. Dashed lines connecting panels $D$ and $F$ show fields
$B_{g}$ for which $g_{\perp }^E / g_0 =1$, and the SO effect is
effectively cancelled. Color legend is in panel $A$. $g$ in panel
$D$ is divided by $10$.} \label{f4}
\end{figure}

Figure \ref{f4} shows that for GaAs both definitions in panels $A$
and $C$ give essentially the same values of $g_{\parallel }$,
where $\Delta E$ is shown in panel $E$. Differences are noticed
only at high magnetic fields for the weakest confinements ($0.7$
and $1.1$ meV). Results are totally different for the
perpendicular field case. Panel $B$ can be understood by looking
at panel $D$ of Fig.\ \ref{f2} (for $1.1$ meV), where it can be
seen that $g_{\perp }^{\sigma _{z}}$ has inverted sign at low
fields, becomes zero for $B_{\perp }$ between $4$ and $8$ T, then
acquires inverted sign again, and suddenly flips back to its
`normal' behavior at $13.7$ T; under even higher fields, $g_{\perp
}^{\sigma _{z}}$ goes to zero since the two lowest level spins are
aligned. Panel $B$ also shows that larger $E_{0}$ values (thus
smaller SO coupling) cancels the field range where $g_{\perp
}^{\sigma _{z}}$ is zero. Notice that the weaker the confinement
the smaller the field where the sign change occurs.

One finds totally different results for $g_{\perp }^{E}$ (panel $D$, with $%
\Delta E$ shown in panel $F$), which may even assume values $12$ times
larger than $g_{0}$ for the smallest $E_{0}$ at low fields; still at low
fields, larger $E_{0}$ tend to reduce $g_{\perp }^{E}$ towards $g_{0}$. At
high fields (inset $G$), $g_{\perp }^{E}$ goes to zero when the level
crossing involving the ground state occurs, and after such field it
increases again. For every $E_{0}$ there is a magnetic field $B_{g}$ --
indicated by the dashed lines connecting panels $D$ and $F$ -- where $%
g_{\perp }^{E}$ goes from higher to smaller values than $g_{0}$. In Panel $F$%
, differently from what occurs in panel $E$ (and in panels $E$ and
$F$ of Fig.\ \ref{f3}), the $B_{g}$-field defines two distinct
phases in the spectrum for a given $E_{0}$: \textit{below}
(\textit{above}) $B_{g}$ the SO coupling \textit{increases}
(\textit{decreases}) the Zeeman sublevel splitting as compared to
the case without SO interaction; one can then say that at
$B_{\perp }=B_{g}$, the SO coupling is \textit{cancelled} by the
magnetic field in the formation of the sublevel splitting. Such
result emphasizes the intricate competition between external
magnetic field and intrinsic SO coupling in QDs.  [Notice that if
a broad Gaussian well or a larger $z_{0}$ is considered, such
phases are not observed.\cite{well}] In GaAs QDs, the anisotropic
nature of the $g$-factor is much more pronounced, despite the
small values of the SO constants.

We conclude with an experimental comparison. Ref.\
[\onlinecite{hanson}] reported Zeeman sublevel splittings in GaAs
QD having $E_{0}=1.1$ meV under in-plane field. They found $\Delta
E\simeq 200$ $\mu $eV at $10$ $T$, while the corresponding curve
in panel $E$ of our Fig.\ \ref{f4} yields $\Delta E\simeq 180$
$\mu $eV. In a linear fit, they found $\left\vert g\right\vert
=0.29\pm 0.01$, and from panel $A$ (panel $C$) of Fig.\ \ref{f4}
one has $\vert g_{\parallel }^{\sigma _{z}}\vert =0.30$ ($\vert%
g_{\parallel }^{E}\vert =0.31$) at $B=10$ T, while $|g_{\parallel%
}|=0.30$ is found at $B=0$ from both definitions.

We have shown how SO coupling is able to tune the electron $g$-factor in QDs
and even change its sign. We have analyzed the interplay between SO and
Zeeman splittings on QD spectra and shown which SO term causes the lowest
ACs in in-plane and perpendicular fields, as well as in different materials.
We have seen that for GaAs QDs under $B_{\perp }$ the ground state has its
spin character \textit{inverted} at low fields if a \textit{narrow} well
confines the system in the $z$-direction. We have identified phases of $%
B_{\perp }$ in the spectrum where the SO interaction increases or decreases
the Zeeman splitting of the lowest QD levels, and explicitly shown the
anisotropic nature of $g$-factor in QDs. All these features would not have
been accessed if a perturbative approach had been used, especially for QDs
with large lateral size.

We thank support from NSF-IMC grant 0336431, CMSS at OU, and the
$21^{st}$ Century Indiana Fund.


\begin{thebibliography}{99}
\bibitem{vicenzo} D. Loss and D. P. DiVincenzo, Phys. Rev. A \textbf{57},
120 (1998).

\bibitem{sallis} G. Salis, Y. Kato, K. Ensslin, D. C. Driscoll, A. C.
Gossard, and D. D. Awschalom, Nature \textbf{414}, 619 (2001).

\bibitem{kato} Y. Kato, R. C. Meyers, D. C. Driscoll, A. C. Gossard, J.
Levy, and D. D. Awschalom, Science \textbf{299}, 1201 (2003).

\bibitem{medeiros} G. Medeiros-Ribeiro, M. V. B. Pinheiro, V. L. Pimentel,
and E. Marega, Appl. Phys. Lett. \textbf{80}, 4229 (2002).

\bibitem{hanson} R. Hanson, B. Witkamp, L. M. K. Vandersypen, L. H. Willems
van Beveren, J. M. Elzerman, and L. P. Kouwenhoven, Phys. Rev. Lett. \textbf{%
91}, 196802 (2003).

\bibitem{lindemann} S. Lindemann, T. Ihn, T. Heinzel, W. Zwerger, K. Ensslin,
K. Maranowski, and A. C. Gossard, Phys. Rev. B \textbf{66}, 195314
(2002).

\bibitem{kiselev} A. A. Kiselev, E. L. Ivchenko, and U. R\"{o}ssler, Phys.
Rev. B \textbf{58}, 16353 (1998).

\bibitem{rodina} A. V. Rodina, Al. L. Efros, and A. Yu. Alekseev, Phys. Rev.
B \textbf{67}, 155312 (2003).

\bibitem{silvio} S. J. Prado, C. Trallero-Giner, A. M. Alcalde, V. L\'{o}%
pez-Richard, and G. E. Marques, Phys. Rev. B \textbf{69}, 201310R (2004).

\bibitem{dest} C. F. Destefani, S. E. Ulloa, and G. E. Marques, Phys. Rev. B
\textbf{69}, 125302 (2004); C. F. Destefani, S. E. Ulloa, and G. E. Marques,
to appear in Phys. Rev. B (10/2004).

\bibitem{raef} E. I. Rashba and Al. L. Efros, Phys. Rev. Lett. \textbf{91},
126405 (2003).

\bibitem{rog} R. de Sousa and S. Das Sarma, Phys. Rev. B \textbf{68}, 155330
(2003).

\bibitem{rashba} Y. A. Bychkov and E. I. Rashba, J. Phys. C \textbf{17},
6039 (1984).

\bibitem{dres} G. Dresselhaus, Phys. Rev. \textbf{100}, 580 (1955).

\bibitem{SOZeeman} M. Val\'{\i}n-Rodr\'{\i}guez, Phys. Rev. B \textbf{70},
033306 (2004).

\bibitem{gover} M. Governale, Phys. Rev. Lett. \textbf{89}, 206802 (2002).

\bibitem{aleiner} I. L. Aleiner and V. I. Fal'ko, Phys. Rev. Lett. \textbf{87%
}, 256801 (2001).

\bibitem{pawel} L. Jacak, A. Wojs, and P. Hawrylak, \textit{Quantum Dots}
(Springer, Berlin, 1998).

\bibitem{parin} Parameters for InSb are: $m=0.014$ m$_{0}$, $g_{0}=-51$, $%
\alpha =500$ \AA $^{2}$, $\gamma =160$ eV\AA $^{3}$, $E_{0}=\hbar \omega
_{0}=15$ meV ($l_{0}=190$ \AA ), $z_{0}=40$ \AA , $dV/dz=-0.5$ meV/\AA .

\bibitem{parga} Parameters for GaAs are: $m=0.067$ m$_{0}$, $g_{0}=-0.44$, $%
\alpha =4.4$ \AA $^{2}$, $\gamma =26$ eV\AA $^{3}$, $E_{0}=\hbar \omega
_{0}=1.1$ meV ($l_{0}=320$ \AA ), $z_{0}=40$ \AA , $dV/dz=-0.5$ meV/\AA .

\bibitem{well} Notice that $\left\langle k_{z}^{2}\right\rangle =(\pi
/z_{0})^{2}$ for a hard wall confinement. If a Gaussian well is considered,
for example, one has $\left\langle k_{z}^{2}\right\rangle =1/z_{0}^{2}$,
which yields a linear BIA term a factor of $10$ smaller than used in this
work, thus decreasing the SO coupling.
\end{thebibliography}
\end{document}